\documentstyle[12pt]{article}
\def\be{\begin{equation}}
\def\ee{\end{equation}}
\def\bea{\begin{eqnarray}}
\def\eea{\end{eqnarray}}
\def\bq{\begin{quote}}
\def\eq{\end{quote}}
\def\bseq{\begin{subequation}}
\def\eseq{\end{subequation}}
\def\bsea{\begin{subeqnarray}}
\def\esea{\end{subeqnarray}}
\def\sbeta{\sin\beta}
\def\cb{\cos\beta}
\def\sa{\sin\alpha}
\def\ca{\cos\alpha}
\def\simlt{\stackrel{<}{{}_\sim}}
\def\simgt{\stackrel{>}{{}_\sim}}
\def\ov{\overline}

\def\epem{e^+e^-}

\def\mh{m_h}
\def\mhh{m_H}
\def\tb{\tan\beta}

\def\mt{m_t}

\def\mw{m_{\rm W}}
\def\mz{m_{\rm Z}}
\def\ma{m_{\rm A}}
\def\matb{(m_A,\tan \beta)}

\def\mhc{m_{H^{\pm}}}
\def\msq{m_{sq}}

\def\msta{m_{\tilde{t}_1}}
\def\mstb{m_{\tilde{t}_2}}

\def\tev{\rm \; TeV}
\def\gev{\rm \; GeV}
\def\fb{\rm \; fb}
\def\pb{\rm \; pb}
\def\sabsq{\sin^2 (\beta - \alpha)}
\def\cabsq{\cos^2 (\beta - \alpha)}
\begin{document}
\begin{titlepage}
\vspace*{-1cm}
\noindent
\phantom{DRAFT 28/3/93}
\hfill{CERN-TH.6792/93}
\vskip 4.0cm
\begin{center}
{\Large\bf SUPERSYMMETRIC HIGGS BOSONS:}
\end{center}
\begin{center}
{\Large\bf A THEORETICAL INTRODUCTION }
\end{center}
\vskip 1.5cm
\begin{center}
{\large F. Zwirner}\footnote{On leave from INFN, Sezione di Padova,
Padua,
Italy.}
\end{center}
\begin{center}
Theory Division, CERN, \\
Geneva, Switzerland \\
\end{center}
\vskip 2.0cm
\begin{abstract}
\noindent
After an introduction to the Higgs sector of supersymmetric
extensions of
the Standard Model, recent results on radiative corrections to Higgs
boson
masses and couplings are reviewed. The phenomenology of
supersymmetric Higgs
searches at large hadron colliders and at a possible linear $\epem$
collider
is also described.
\end{abstract}
\vskip 2.0cm
\begin{center}
{\em
Invited talk at the Workshop `Ten years of SUSY confronting
experiment',
\\
CERN, 7--9 September 1992, to appear in the Proceedings
}
\end{center}
\vfill{
CERN-TH.6792/93
\newline
\noindent
February 1993}
\end{titlepage}
\setcounter{footnote}{0}
\section{The Higgs sector of SUSY models}

It has been known for a long time [\ref{fayet}] that realistic
supersymmetric
extensions of the Standard Model (SM) require at least two Higgs
doublets.
Higgs fields must be embedded in chiral supermultiplets, whose
physical
degrees of freedom are a complex spin-$0$ boson and a two-component
spin-$1/2$ fermion, and their Yukawa couplings to the matter fermions
are
encoded in the superpotential, an analytic function of the chiral
superfields\footnote{For a pedagogical review of the formalism of
$N=1$ global
supersymmetry, see e.g. ref.~[\ref{sohnius}].}. At least two Higgs
doublets
are then needed: 1) to give masses to all quarks and charged leptons
via
superpotential couplings; 2) to avoid higgsino-induced chiral
anomalies (in
the absence of other anomaly-cancellation mechanisms); 3) to avoid
the
existence of a massless charged spin-$1/2$ particle in the
gaugino-higgsino
sector.

\subsection{The minimal model}

In the Minimal Supersymmetric extension of the Standard Model (MSSM),
the Higgs sector consists of just two doublets:
\be
H_1
\equiv
\left(
\begin{array}{c}
H_1^0
\\
H_1^-
\end{array}
\right)
\sim
(1,2,-1/2)
\, ,
\;\;\;\;\;
H_2
\equiv
\left(
\begin{array}{c}
H_2^+
\\
H_2^0
\end{array}
\right)
\sim
(1,2,+1/2)
\, ,
\ee
and, after imposing $R$-parity conservation, the superpotential reads
\be
\label{superpotential}
w = h^U Q U^c H_2 + h^D Q D^c H_1
+ h^E L E^c H_1 + \mu H_1 H_2 \, .
\ee
The first three addends in (\ref{superpotential}) originate an
acceptable
set of Yukawa couplings, with automatic flavour conservation in
tree-level
neutral currents. The last one, containing the mass parameter $\mu$,
corresponds to a globally supersymmetric Higgs mass term, to be
discussed
in more detail later. After inclusion of general soft
supersymmetry-breaking
terms, which parametrize our ignorance about the mechanism for
supersymmetry
breaking in the underlying fundamental theory, the tree-level
potential
of the MSSM reads
\bea
\label{v0}
V_0 & = &  m_1^2 \left| H_1 \right|^2 + m_2^2 \left| H_2
\right|^2 + m_3^2 \left( H_1 H_2 + {\rm h.c.} \right) \nonumber \\
& + &  {g^2 \over 8} \left( H_2^{\dagger} {\vec\sigma} H_2
+ H_1^{\dagger} {\vec\sigma} H_1 \right)^2 +
{g' \, ^2 \over 8} \left(  \left| H_2 \right|^2 -
\left| H_1 \right|^2 \right)^2 \, ,
\eea
where $m_1^2,m_2^2,m_3^2$ are essentially arbitrary mass parameters,
$g$ and $g'$ are the $SU(2)_L$ and $U(1)_Y$ coupling constants,
respectively, and ${\vec\sigma}$ are the Pauli matrices.
A crucial difference between the Higgs potentials of the
SM and of the MSSM is evident from (\ref{v0}): in the SM
the quartic scalar coupling is determined by an arbitrary
parameter $\lambda$, proportional to the SM Higgs mass $m_{\varphi}$.
In (\ref{v0}), despite the presence of two Higgs doublets, the
quartic
scalar couplings do not contain any arbitrary parameter, but are
determined
by the $SU(2)_L \times U(1)_Y$ gauge couplings: supersymmetry
strongly
constrains the form of the scalar potential.

To discuss the mass spectrum, it is not restrictive to consider the
following expansion
\be
H_1
=
\left(
\begin{array}{c}
{\displaystyle
v_1 + {S_1 + i P_1 \over \sqrt{2}}}
\\
H_1^-
\end{array}
\right)
\, ,
\;\;\;\;\;\;
H_2
=
\left(
\begin{array}{c}
H_2^+
\\
{\displaystyle
v_2 + {S_2 + i P_2 \over \sqrt{2}}}
\end{array}
\right)
\, ,
\ee
and to choose $m_3^2$ real and negative, so that the vacuum
expectation values $v_1 \equiv \langle H_1^0 \rangle$ and $v_2 \equiv
\langle H_2^0 \rangle$ can be taken to be real and positive.

One can then read from the MSSM Lagrangian the different mass terms
in
the $R$-even sector of the theory. The weak boson masses are given by
\be
m_W^2 = {g^2 \over 2} (v_1^2 + v_2^2) \, ,
\;\;\;\;\;
m_Z^2 = {g^2 + g'^2 \over 2} (v_1^2 + v_2^2) \, ,
\ee
and the fermion mass matrices by
\be
m^U = h^U v_2 \, ,
\;\;\;\;\;
m^D = h^D v_1 \, ,
\;\;\;\;\;
m^E = h^E v_1 \, .
\ee
A physical constraint comes from the fact that the combination
$(v_1^2+v_2^2)$, which determines the $W$ and $Z$ boson masses, must
reproduce their measured values. Once this constraint is imposed, at
the classical level the MSSM Higgs sector contains only two
independent
parameters: they can be conveniently taken to be $\tb \equiv v_2/v_1$
and one combination of the mass parameters (of the three original
ones,
two are removed by the minimization conditions).

The spin-0 boson mass matrices factorize into $2 \times 2$ blocks,
corresponding to the charged, neutral CP-odd and neutral
CP-even sectors. One can easily identify the Goldstone bosons
$G^+ = - \cb (H_1^-)^* + \sbeta H_2^+$, $G^- = (G^+)^*$ and $G^0
= - \cb P_1 + \sbeta P_2$. The orthogonal combinations $H^+ = \sbeta
(H_1^-)^* + \cb H_2^+$, $H^- = (H^+)^*$ and $A = \sbeta P_1 + \cb
P_2$ correspond to physical particles, with masses
\be
m_A^2 = - m_3^2 \left( \tb + \cot \beta \right) \, ,
\ee
and
\be
\mhc^2 = \mw^2 + \ma^2 \, .
\ee
A convenient choice, which will be adopted here, is to take as
independent
parameters $m_A$ and $\tb$. The mass matrix of the neutral CP-even
sector
then reads
\be
\label{cpeven}
\left( {\cal M}_R^0 \right)^2 =
\left[
\pmatrix{
\cot\beta & -1 \cr
-1 & \tan\beta \cr}
{m_Z^2 \over 2}
+
\pmatrix{
\tan\beta & -1 \cr
-1 & \cot\beta \cr}
{m_A^2 \over 2}
\right]
\sin 2\beta
\, .
\ee
Defining the mass eigenstates as
\be
h = - \sa \, S_1 + \ca \, S_2 \, ,
\;\;\;\;\;
H = \ca \, S_1 + \sa \, S_2 \, ,
\ee
one obtains
\be
m_{h,H}^2 = {1 \over 2} \left[
m_A^2 + m_Z^2 \mp \sqrt{(m_A^2 + m_Z^2)^2
- 4 m_A^2 m_Z^2 \cos^2 2 \beta}
\right],
\ee
and also celebrated inequalities such as $m_W, m_A < m_{H^{\pm}}$,
$m_h
<  m_Z |\cos 2 \beta | < m_Z < m_H$, $m_h <  m_A < m_H$. Similarly,
one
can easily compute all the Higgs boson couplings by observing that
the
mixing angle $\alpha$, required to diagonalize the mass matrix
(\ref{cpeven}), is given by
\be
\cos 2 \alpha = - \cos 2 \beta \; { { m_A^2  - m_Z^2 }\over {
 m_H^2 - m_h^2  }} \ ,\ \ - {\pi \over 2} < \alpha\  {\leq}\  0.
\ee
For example, the couplings of the three neutral Higgs bosons to
vector
boson and fermion pairs are easily obtained from the SM Higgs
couplings: it
is sufficient to multiply the latter by the $\alpha$- and
$\beta$-dependent
factors summarized in Table~1.
\begin{table}
\begin{center}
\begin{tabular}{|c|c|c|c|}
\hline
& & & \\
&
$
\begin{array}{c}
d\overline{d},s\overline{s},b\overline{b} \\
e^+e^-,\mu^+\mu^-,\tau^+\tau^-
\end{array}
$
& $u\overline{u},c\overline{c},t\overline{t}$
& $W^+W^-,ZZ$ \\
& & & \\
\hline
& & & \\
$h$
& $- \sin \alpha / \cos \beta $
& $  \cos \alpha / \sin \beta $
& $  \sin \, (\beta -\alpha)  $ \\
& & & \\
\hline
& & & \\
$H$
& $\cos \alpha / \cos \beta$
& $\sin \alpha / \sin \beta$
& $\cos \, (\beta -\alpha)  $
\\
& & & \\
\hline
& & & \\
$A$
& $-i\gamma_5 \tan \beta $
& $-i\gamma_5 \cot \beta $
& $0$
\\
& & & \\
\hline
\end{tabular}
\caption{Correction factors for the couplings of the MSSM neutral
Higgs
bosons to fermion and vector boson pairs.}
\label{couplings}
\end{center}
\end{table}
The remaining tree-level Higgs boson couplings in the MSSM can be
easily
computed and are summarized, for example, in ref.~[\ref{hunter}].

An important consequence of the tree-level structure of the Higgs
potential
is the existence of at least one neutral CP-even Higgs boson with
mass smaller
than $m_Z$ ($h$) or very close to it ($H$), and significantly coupled
with the
$Z$ boson. In the past, this raised the hope that the crucial
experiment
on the MSSM Higgs sector could be entirely performed at LEP~II (with
sufficient centre-of-mass energy, luminosity and $b$-tagging
efficiency),
and took some interest away from higher energy colliders. However, it
was
recently pointed out [\ref{pioneer}] that the Higgs-boson masses are
subject
to large, finite radiative corrections, dominated by loops involving
the top
quark and its supersymmetric partners. We shall summarize
the present theoretical status of these corrections in section~2, and
their
implications for future colliders in section~3 (the phenomenology of
SUSY Higgs
bosons at LEP~I and LEP~II is discussed in other talks at this
Workshop
[\ref{lep1},\ref{lep2}]). We would now like to conclude this
section with a few comments on non-minimal supersymmetric Higgs
sectors, which
might be useful to understand the relevance of the minimal case.

\subsection{Non-minimal models}

As in non-supersymmetric model building, also in the presence of
low-energy
supersymmetry one can abandon the economy principle, adding further
chiral
superfields to the MSSM Higgs sector. Some of these additions can
create severe phenomenological problems, which can be solved only at
the
price of rather artificial constructions. Extra doublets, for
example,
are potentially dangerous sources of tree-level flavour-changing
neutral
currents and charge-breaking minima. Extra Higgses in
higher-dimensional
representations of $SU(2)_L$ are difficult to reconcile with the
measured
value of $m_W/m_Z$; moreover, they do not seem to appear in the
massless
spectrum of realistic four-dimensional string constructions. A less
dangerous option is to introduce one extra singlet (or more) under
$SU(2)_L
\times U(1)_Y$, as in the first models by Fayet [\ref{fayet}]. A
mildly
attractive possibility along these lines is the introduction of just
one
extra singlet superfield, $N$, with a purely cubic superpotential
[\ref{mnmssm}]
\be
\mu H_1 H_2 \longrightarrow \lambda H_1 H_2 N + k N^3 \, ,
\ee
and a corresponding modification in the associated soft SUSY-breaking
part of the scalar potential
\be
m_3^2 \left ( H_1 H_2 + {\rm h.c.} \right)
\longrightarrow
\left( \lambda A_{\lambda} H_1 H_2 N + k A_k N^3 + {\rm h.c.} \right)
\, .
\ee
{}From the point of view of the low-energy effective theory, this
`minimal-non-minimal' model contains two more parameters and two more
neutral
states than the MSSM. The role of the mixing masses $\mu$ and
$m_3^2$,
necessary to obtain an acceptable breaking of the electroweak gauge
symmetry,
is effectively played by the quantities $\lambda x$ and $\lambda
A_{\lambda}
x$, where $x \equiv \langle N \rangle$. The MSSM is recovered by
taking the
limit $x \to \infty$, while keeping $\lambda x$ and $k x$ fixed, in
which case
the two additional neutral spin-0 states become superheavy and
decouple from
the low-energy theory. In general, however, this model has a much
more
complicated phenomenology [\ref{gang}] than the MSSM. Before using it
as an
alternative paradigm for low-energy supersymmetry, it might be useful
to
carefully analyse its motivations.

One of the original motivations for the minimal-non-minimal model is
the
so-called $\mu$-problem [\ref{muproblem}] of the MSSM. From the point
of view
of the low-energy effective theory (the MSSM), the superpotential
mass
parameter $\mu$ is not related to the scale of supersymmetry
breaking. On the
other hand, $\mu = m_3 = 0$ would give an unacceptable axion, and in
any case
$\mu=0$ is excluded by the present LEP data [\ref{lepsusy}]. In the
absence of
a theoretical explanation for its existence, a supersymmetric mass
term at the
electroweak scale is clearly unsatisfactory.  A second, more recent
motivation
is the observation that supergravity models describing the low-energy
limit of
four-dimensional string constructions have only cubic (or
higher-dimensional)
couplings  among the light fields with non-trivial gauge quantum
numbers. Two
problems then have to be solved in the fundamental theory giving the
MSSM in
the low-energy limit: 1) Why is $\mu=0$ (instead of $\mu=M$, where
$M$ is some
superheavy scale) in the limit of unbroken supersymmetry? 2) How can
one
generate a non-vanishing $\mu$,  of the order of the electroweak
scale, after
the breaking of local supersymmetry?

As for the first problem, which is obviously present also in
non-minimal
models, one could simply argue that, if there is no supersymmetric
mass term of
this kind to begin with, non-renormalization theorems protect it from
large
radiative corrections. More ambitious suggestions to explain this
fact resort
to the so-called `missing-partner'
mechanism [\ref{missing}], or to the idea of seeing the Higgs
doublets as
pseudo-Goldstone bosons,
associated with the quotient of some large global symmetry group over
the
grand-unification group
[\ref{pseudo}].  As for the second problem, there are possible
solutions that
do without additional
Higgs singlets at low energy. For example, one can start from a
supergravity
theory with a purely cubic
superpotential in the observable sector and, if the geometrical
structure of
the theory is appropriate,
generate [\ref{giudice}] a non-vanishing $\mu$-term, proportional to
the
gravitino mass $m_{3/2}$, in
the low-energy effective theory with softly broken global
supersymmetry,
obtained by taking the flat
limit [\ref{flat}] $M_P \to \infty$. Alternatively, one can think of
possible
non-renormalizable
superpotential terms of the form $w \ni \phi^n H_1 H_2 / M^{n-1}$,
where $M$ is
some very large scale
and the symbol $\phi$ stands for singlet fields getting a vacuum
expectation
value $\langle \phi
\rangle \sim \tilde{M} < M$: a globally supersymmetric mass $\mu \sim
\tilde{M}^n / M^{n-1}$ is then
generated in the low-energy theory [\ref{muproblem}].

{}From the previous considerations, it should be apparent that the
solution of
the $\mu$-problem does
not necessarily imply an extension of the MSSM Higgs sector at the
level of the
low-energy effective
theory. Moreover, the minimal-non-minimal model has also some
potential
drawbacks. First of all,
models with singlets coupled with the superheavy sector of the theory
might
develop dangerous
instabilities along the singlet direction [\ref{sliding}].
Furthermore, in the
minimal-non-minimal
model it must be true that $k \ne 0$ to avoid a global axionic
symmetry,
spontaneously
broken by the
expectation values of the Higgs fields. On the other hand, $k \ne 0$
seems
difficult to obtain in the
low-energy limit of four-dimensional string models: singlets under
the standard
model gauge group are
in general charged under some gauge group broken at high energy, and
this is
sufficient to forbid a
purely cubic coupling in the superpotential.

In conclusion, non-minimal models are certainly not excluded by the
present
theoretical and experimental knowledge, but for the moment they do
not appear
to have stronger motivations than the MSSM.

\section{Radiative corrections to SUSY Higgs masses and couplings}

Radiative corrections to the parameters of the Higgs boson sector in
the MSSM
have recently received much attention. After the discovery
[\ref{pioneer}] that
top and stop loops can cause large corrections to the masses of the
neutral
CP-even Higgs bosons, radiative corrections to Higgs boson masses and
couplings
have been computed by a variety of methods: the renormalization group
approach
[\ref{rga}], the effective potential approach [\ref{epa}], and the
diagrammatic
approach [\ref{da},\ref{bz}].

The renormalization-group approach assumes that there are two (or
more)
widely separated mass scales, for example
\be
\label{brute}
M_{SUSY} \; (\sim \msta \sim \mstb \sim \ldots \sim \mhh \sim \mhc
\sim \ma) \gg \mz \; (\sim \mh \sim \mt) \, ,
\ee
and considers the effective theory for the degrees of freedom lighter
than $M_{SUSY}$. It then solves (non-supersymmetric) renormalization
group equations to obtain running parameters down to the scale $Q =
\mz$,
imposing the tree-level relations of the MSSM as boundary conditions
at
the scale $Q = M_{SUSY}$. This approach has the
advantage of resumming the leading corrections, proportional to
$\log(M_{SUSY}/\mz)$, so that even the case of $M_{SUSY}$ orders of
magnitude larger than $\mz$ can be dealt with in perturbation theory.
On the other hand, if supersymmetry is to solve the naturalness
problem
of the Standard Model, one expects the various mass parameters of the
MSSM
to be scattered around the electroweak scale, $G_F^{-1/2} \simeq 250
\gev$,
so that assumption (\ref{brute}) breaks down.

The effective-potential approach consists in identifying the Higgs
boson
masses and self-couplings with the corresponding derivatives of the
one-loop effective potential, evaluated at the minimum. By
definition,
this approach evaluates all Higgs self-energies and vertices at
vanishing
external momentum. In the case of radiative corrections to Higgs
boson
masses, this was shown to be a rather accurate
approximation\footnote{
Actually, when the external momentum (i.e. the Higgs mass) approaches
or
exceeds the threshold of the internal particles, the full correction
can be
rather different from the zero-momentum one. However, in that case
corrections
themselves are small, either in the absolute sense or relatively to
the
(increased) tree-level mass.}. Other possible drawbacks of the
effective
potential approach are the gauge- and scale-dependence of the
associated
quantities. These are not serious problems in the computation of the
mass
corrections: the dominant ones come from quark and squark loops,
which
introduce no spurious dependences on the gauge parameter into the
results;
also, wave-function renormalization effects, responsible for the
scale
dependence, are generally small with respect to the overall mass
corrections.

The diagrammatic approach consists in performing the complete
one-loop renormalization programme, specifying unambiguously the
input
parameters and the relations between renormalized parameters and
physical
quantities. This approach gives the most precise computational
tool in the case of supersymmetric particle masses spread around the
electroweak scale, and results that are formally gauge- and
scale-independent. Since corrections can be numerically large,
however,
one has to pay attention and conveniently  improve the na\"{\i}ve
one-loop
calculations when necessary.

To simplify the discussion, in the following we shall take a
universal soft
supersymmetry-breaking squark mass, $\msq$, and assume negligible
mixing in the
stop mass matrix, $A_t = \mu = 0$. More complete formulae for
arbitrary values
of the parameters are available, but the qualitative features
corresponding to
the above choices are representative of a very large region of
parameter space.
In the case under consideration, and working at leading order in the
top-quark
Yukawa coupling, the neutral CP-even mass matrix is modified at one
loop as
follows
\be
{\cal M}_R^2 = \left( {\cal M}_R^0 \right)^2
+
\Delta {\cal M}_R^2 \, ,
\ee
where
\be
\label{cpeven1}
\left( \Delta {\cal M}_R^2 \right)_{11,12,21}
=
0 \, ,
\;\;\;\;\;
\left( \Delta {\cal M}_R^2 \right)_{22}
=
{3 \over 8 \pi^2} {g^2 m_t^4 \over m_W^2 \sin^2 \beta} \log
\left( 1 + {\msq^2 \over m_t^2} \right) \, .
\ee
It is then a simple exercise to derive the one-loop-corrected
eigenvalues
$m_h$ and $m_H$, as well as the mixing angle $\alpha$ associated with
the
one-loop-corrected mass matrix ${\cal M}_R^2$. The most striking fact
in
eq.~(\ref{cpeven1}) is that the correction $( \Delta {\cal M}_R^2
)_{22}$
is proportional to $(m_t^4/m_W^2)$ for fixed $(\msq/m_t)$. This
implies that,
for $m_t$ in the presently allowed range, the tree-level predictions
for $m_h$
and $m_H$ can be badly violated, as for the related inequalities. The
other
free parameter in eq.~(\ref{cpeven1}) is $\msq$, but the dependence
on it is
much milder.

The above formulae have been generalized to arbitrary values of the
parameters
in the stop mass matrix, and the effects of other virtual particles
in the
loops have been included. Renormalization-group methods have been
used
to resum the large logarithms that arise when the typical scale of
supersymmetric particle masses, $M_{\rm SUSY}$, is much larger than
$m_Z$.
Two-loop corrections have been computed in the leading logarithmic
approximation, and found to be small. After all these refinements,
eq.~(\ref{cpeven1}) still gives the most important mass correction in
the most plausible region of parameter space.

To illustrate the impact of eq.~(\ref{cpeven1}), we display in fig.~1
contours
of $m_h^{max}$ (the maximum value of $m_h$, reached for $m_A \gg m_Z$
and $\tb
\gg 1$), in the $(m_t,\msq)$ plane. The calculation has been
performed in the
effective potential approach, including top, bottom, stop, sbottom
loops, and
neglecting mixing in the stop and sbottom mass matrices. For very
large
values of $m_t$ and $\msq$, the renormalization-group improvement can
lower
the actual value of $m_h^{max}$ by a non-negligible amount. In the
following, when making numerical examples we shall choose the
numerical
values $m_t = 140 \gev$, $\msq = 1 \tev$: for this parameter choice,
the
effect of the renormalization-group improvement is of the order of a
few GeV,
comparable with other residual theoretical uncertainties.

The computation of radiative corrections can be extended to the other
parameters of the MSSM Higgs sector. For example, one-loop
corrections to the
charged Higgs mass have been computed and found to be small, at most
a few GeV,
for generic values of the parameters.

Whilst radiative corrections to Higgs boson masses are by now well
under
control, the study of radiative corrections to Higgs boson couplings
is
still at a less refined stage. In the case of Higgs boson
self-couplings,
which control decays such as $H \to hh$, $H \to AA$ and $h \to AA$
when they
are kinematically allowed, radiative corrections can be numerically
large.
Detailed computations of these corrections have been performed by a
variety
of methods. For example, the leading radiative correction to the
cubic $Hhh$
coupling can be written as $\lambda_{Hhh} = \lambda_{Hhh}^0 + \Delta
\lambda_{Hhh}$, where
$$
\lambda_{Hhh}^0
=
- {i g m_Z \over {2 \cos \theta_W}}
[ 2 \sin (\beta + \alpha) \sin 2 \alpha
- \cos (\beta + \alpha) \cos 2 \alpha]
\eqno (2.9)
$$
is the tree-level coupling, and
$$
\Delta \lambda_{Hhh}
=
- {i g m_Z \over {2 \cos \theta_W}}
  {3 g^2 \cos^2 \theta_W \over {8 \pi^2}}
  {\cos^2  \alpha \sin \alpha \over  {\sin^3 \beta}}
  {m_t^4 \over m_W^4}
  \left( 3 \log   {\msq^2 + m_t^2 \over m_t^2}
       - 2 {\msq^2 \over \msq^2 + m_t^2} \right).
\eqno (2.10)
$$
Notice the explicit dependence on the ratio $(m_t/m_W)^4$. Given the
fact that,
in addition to the masses of the virtual particles in the one-loop
diagrams,
two different mass scales are involved in the decays $H \to hh$, $H
\to AA$
and $h \to AA$, the mass of the decaying particle and the mass of the
decay
products, one might suspect that momentum-dependent effects, which
are
neglected in the
renormalization-group and in the effective-potential approaches,
could play a
role. This
problem was recently studied in ref.~[\ref{bz}], which performed a
full
diagrammatic computation of
the decay rate $\Gamma(H \to hh)$, including top, bottom, stop and
sbottom
loops. A typical result is
shown in fig.~2, which gives $\Gamma(H \to hh)$ as a function of
$m_H$ for a
representative parameter
choice. One finds that, for $\tb$ close to 1, there can be very large
corrections to the `improved
tree-level' approximation, which amounts to plugging the one-loop
corrected
value of the mixing
angle $\alpha$ into the tree-level formulae. Also, for $\tb$ close to
1 and
$m_H \simgt 2 m_t$ the
full diagrammatic result can significantly differ from the one
obtained in the
effective potential
approach. The latter method, however, remains a good aproximation in
the region
of parameter space
which is most relevant for $H$ searches at future colliders.

One should also consider radiative corrections to Higgs couplings to
vector
boson and fermions. In most phenomenological studies, they have been
taken into
account only approximately, by improving the tree-level formulae with
one-loop
corrected values of the $H$--$h$ mixing angle, $\alpha$, and with
running
fermion masses, evaluated at the typical scale $Q$ of the process
under
consideration. Residual corrections are expected to be numerically
small in the
experimentally interesting regions of parameter space, as recently
verified on
a number of explicit examples.

Finally, it is interesting to ask what happens to the upper
bound on the lightest Higgs mass if one goes from the MSSM to
non-minimal models. From the point of view of the low-energy
effective theory, the bound is no longer valid as long as
supersymmetry and the particle content allow for some arbitrary
quartic coupling in the tree-level potential. In the
minimal-non-minimal model,
for example, the tree-level bound is modified
into
\be
m_h^2 \le m_Z^2 \left( \cos^2  2 \beta + {2 \lambda^2 \cos^2
\theta_W \over g^2} \sin^2 2 \beta \right) \, .
\ee
Large values of the arbitrary coupling $\lambda$ can give large
violations of the MSSM bound already at the tree level. On the other
hand, a bound on the value of $\lambda$ at the electroweak scale, and
thus on $m_h$, can be obtained by requiring that the running coupling
constants of the model, including the Higgs self-coupling and the
Yukawa couplings, remain perturbative up to the grand-unification
scale [\ref{landau}]. This argument can be put forward also in
non-supersymmetric models, including the SM [\ref{cmpp}]; but, in
supersymmetric models it is particularly motivated by the success of
supersymmetric grand unification.  The existence of effective
infrared fixed
points for the top Yukawa coupling $h_t$ and the Higgs self-coupling
$\lambda$
make this bound particularly stringent. Of course, one has to add the
finite radiative corrections due to SUSY-breaking effects, as in the
MSSM. In the case of the minimal-non-minimal model, the result is
displayed in fig.~3, assuming a universal soft SUSY-breaking mass
$M_{\rm SUSY}
= 1 \tev$.  One can see that the absolute upper bound on the lightest
neutral
Higgs boson is of the order of 140 GeV, only slightly higher than the
corresponding bound in the MSSM.

\section{SUSY-Higgs searches at future colliders}

The relevant processes for MSSM Higgs boson searches at LEP are
$\epem \to
Z^* \to h Z^*$ and $\epem \to Z^* \to h A$, which play a
complementary role,
since their rates are proportional to $\sabsq$ and $\cabsq$,
respectively.
Updated experimental limits that take radiative corrections into
account have
been presented at this Workshop by Grivaz [\ref{lep1}]. Neglecting
mixing
effects in the stop sector, and varying $m_t$ and $\msq$ over
plausible ranges,
one gets $m_h > 43 \gev$, $m_A > 21 \gev$. More stringent limits can
be
obtained for specific choices of $m_t$ and $\msq$.

The context in which the impact of radiative corrections is most
dramatic is the search for MSSM Higgs bosons at LEP~II. A detailed
evaluation
of the LEP~II discovery potential can be made only if crucial
theoretical
parameters, such as the top-quark mass and the various soft
supersymmetry-breaking masses, and experimental parameters, such as
the
centre-of-mass energy, the luminosity, and the $b$-tagging
efficiency, are
specified.  An updated assessment of the LEP~II discovery potential
has been
presented at this Workshop by Treille [\ref{lep2}], and further
details can
be found in a recent study by Janot [\ref{janot}]. The bottomline is
that,
with the planned machine parameters, LEP~II will not be sensitive to
some
regions of the parameter space characterizing the SUSY Higgs sector,
even in
the most restrictive case of the MSSM. Of course, one should keep in
mind that
there is, at least in principle, the possibility of further extending
the
maximum LEP energy up to values as high as $\sqrt{s} \simeq
230$--$240 \gev$,
at the price of introducing more (and more performing)
superconducting cavities
into the LEP tunnel. This could allow a probing of the most plausible
region of
the parameter space of the MSSM and of its extensions, up to Higgs
mass values
$m_h \simlt 130$--$140 \gev$.

\subsection{The LHC and the SSC}

A natural question to ask is whether, assuming completion of the
LEP~II project
with the foreseen parameters, the LHC and the SSC can explore the
full
parameter space of the MSSM Higgs bosons. A systematic study of this
problem,
including radiative corrections, has recently been started in
refs.~[\ref{kz},\ref{others}].
The analysis is complicated by the fact that the $R$-odd particles
could play a
role both in the production (via loop diagrams) and in the decay (via
loop
diagrams and as final states) of the MSSM Higgs bosons. For
simplicity, one can
concentrate on the most conservative case, in which all $R$-odd
particles are
heavy enough not to play any significant role. Still, one needs to
perform a
separate analysis for each $(m_A,\tb)$ point, to include radiative
corrections
(depending on additional parameters such as $m_t$ and $\msq$), and to
consider
Higgs boson decays involving other Higgs bosons. We make here only a
few
general remarks on the LHC case, for our representative parameter
choice,
sending the interested reader to refs.~[\ref{kz},\ref{others}] for a
more
complete discussion and recent simulation work.

Beginning with the neutral states, when $h$ or $H$ are in the
intermediate mass range (80--130 GeV) and have approximately SM
couplings, the best prospects for detection are offered, as in the
SM, by their $\gamma \gamma$ decay mode. In general, however,
$\sigma \cdot BR (h,H \to \gamma \gamma)$ is smaller than for a
SM Higgs boson of the same mass. As a rather optimistic estimate
of the possible LHC sensitivity, we display in fig.~4 lines in
the $\matb$ plane corresponding to $\sigma \cdot BR (h,H \to \gamma
\gamma) \sim 30 \fb$. The contour line for $h$ is shown only for
$m_h \simgt 80 \gev$. Only for $m_A \simgt 200 \gev$, $\tb \simgt 3$
(in the case of $h$) and in the shaded area (in the case of $H$),
does the
$\gamma \gamma$ signal exceed the chosen reference value. Very
similar
considerations can be made for the production of $h$ or $H$, decaying
into $\gamma \gamma$, in association with a $W$ boson or with a $t
\ov{t}$
pair, giving an extra isolated lepton in the final state. When $H$
and $A$
are heavy, in general one cannot rely on the $ZZ \to 4 l^{\pm}$
$(l=e,\mu)$
decay mode, which gives the `gold-plated' Higgs signature in the SM
case,
since $H$ and $A$ couplings to vector-boson pairs are strongly
suppressed:
only for small $\tan \beta$ and $150 \; {\rm GeV} \simlt m_H \simlt 2
m_t$
might the decay mode $H \rightarrow Z Z \rightarrow 4 l^{\pm}$ still
be viable
despite the suppressed branching ratio. Taking into account that with
sufficient experimental resolution one could exploit the small $H$
width, as an
estimate
of the possible LHC sensitivity we take $\sigma \cdot BR (H \to 4
l^{\pm})
\sim 1 \fb$ ($l=e,\mu$). This contour defines the area in fig.~4
indicated by
the label $H \to 4 l$. For very large values of $\tan \beta$, and
moderately
large $m_A$, one could take advantage of the enhanced production
cross-sections
and of the unsuppressed decays into $\tau^+ \tau^-$ to obtain a
visible signal
for one or more of the MSSM neutral Higgs bosons, and in particular
for $H$
and $A$, whose masses can be significantly larger than 100 GeV. The
simulation
work for this process is still at a rather early stage, so that no
definite
conclusion can be drawn yet. For reference, the dotted line in fig.~4
corresponds to a (somewhat arbitrary) interpolation of $\sigma \cdot
BR
(A,H \to \tau^+ \tau^-) \sim 10 \pb$ at $m_{H,A} \sim 100 \gev$ and
$\sigma \cdot BR (A,H \to \tau^+ \tau^-) \sim 1 \pb$ at $m_{H,A} \sim
400 \gev$.

Finally, in the region of parameter space corresponding to $m_A
\simlt m_Z$,
the charged Higgs could be discovered via the decay chain $t \to b
H^+ \to
b \tau^+ \nu_{\tau}$, which competes with the standard channel $t \to
b W^+
\to b l^{\pm} \nu_l$ ($l=e,\mu,\tau$). A convenient parameter is the
ratio
$R \equiv BR(t \to \tau^+ \nu_{\tau} b)/BR(t \to \mu^+ \nu_{\mu} b)$,
which
measures the violation of lepton universality in top decays. As an
estimate
of the LHC sensitivity, we take $R > 1.15$. The corresponding region
of the
$(m_A,\tb)$ plane is indicated by the label $H^{\pm} \to \tau \nu$ in
fig.~4.

For all processes considered above, similar remarks apply also to the
SSC.
For reference we also show, as dashed lines in fig.~4, contours
associated with
two benchmark values of the total cross-section $\sigma ( e^+ e^-
\rightarrow
hZ,HZ,hA,HA)$, which should give a rough measure of the LEP~II
sensitivity. The
lower line corresponds to $\sigma = 0.2 \pb$ at $\sqrt{s} = 175
\gev$, which
could be seen as a rather conservative estimate of the LEP~II
sensitivity.
The upper line corresponds to $\sigma = 0.05 \pb$ at $\sqrt{s} = 190
\gev$,
which could be seen as a rather optimistic estimate of the LEP~II
sensitivity.

In summary, a global look at fig.~4 shows that there is a high degree
of
complementarity between the regions of parameter space accessible to
LEP~II
and to the LHC/SSC. However, for our representative choice of
parameters,
there is a non-negligible region of the $(m_A,\tb)$ plane that is
presumably
beyond the reach of LEP~II and of the LHC/SSC. This potential problem
could
be solved, as we said before, by a further increase of the LEP~II
energy
beyond the reference value of $\sqrt{s} \simlt 190 \gev$. Otherwise,
one might
need a higher-energy $\epem$ collider, for a full exploration of the
parameter
space describing the MSSM Higgs sector. On the other hand, one should
not
forget that another important test of the MSSM will be provided by
squark and
gluino searches at the LHC/SSC, which should be sensitive to most of
the
theoretically motivated parameter space. One should also keep in mind
that
indirect information on the particle spectrum of the MSSM, including
its
extended Higgs sector, could come from lower-energy precision data.
The possible effects of virtual supersymmetric particles on LEP
observables
have already been mentioned at this Workshop [\ref{virtual}]. Another
interesting effect, recently re-emphasized in refs.~[\ref{bsga}], is
the
contribution of the charged-Higgs loop to the rare decay $b \to s
\gamma$,
which in the SM proceeds via a $W$-boson loop. The theoretical and
experimental
errors on the inclusive radiative $B$-decay could already be small
enough to
put non-trivial constraints on the particle spectrum of the MSSM. In
particular, in the limit of very heavy $R$-odd particles one could
identify an
excluded region in the $\matb$ plane, corresponding to low values of
$m_{H^\pm}$: the precise form of such a region strongly depends on
the assumed
theoretical uncertainties. Furthermore, loops of (relatively light)
virtual
supersymmetric particles can give rise to partial cancellations with
the $W$
and Higgs loops, thus allowing for values of the decay rate on both
sides of
the SM prediction [\ref{bsgb}].

\subsection{High-energy linear $\epem$ colliders}

We now review, following ref.~[\ref{ee500}], the main production
mechanisms of
neutral SUSY Higgses in $e^+e^-$ collisions at very high energy, say
$\sqrt{s}
= 500$ GeV, namely:
\be
\label{processes}
\begin{array}{ll}
e^+e^- \rightarrow h Z \, , H A \, , h \nu \ov{\nu} \, , h e^+ e^-
& [\sigma \propto \sin^2(\beta-\alpha)] \, ,
\\
e^+e^- \rightarrow H Z \, , h A \, , H \nu \ov{\nu} \, , H e^+ e^-
& [\sigma \propto \cos^2(\beta-\alpha)] \, .
\end{array}
\ee
Other production mechanisms of interest are discussed in
refs.~[\ref{ee500},\ref{othersee}], and details about experimental
searches
can be found in refs.~[\ref{janot}]. It is useful to roughly estimate
the
cross-section for which we believe that any of the listed processes
will be
detectable. A cross-section of $0.01$~pb will lead to 25 events for
an
integrated luminosity of $10$~fb$^{-1}$ after multiplying by an
efficiency of
25\%: the latter is a crude estimate of the impact of detector
efficiencies,
cuts, and branching ratios to usable decay channels. One keep this
benchmark
cross-section value in mind as a rough criterion for where in
parameter space a
particular reaction can be useful.

Figure~5 shows contours of $\sigma (  e^+e^- \rightarrow h Z)$ and
$\sigma (
e^+e^- \rightarrow H Z)$ in the $(m_A,\tan\beta)$ plane. Owing
to the much higher energy with respect to the standard LEP~II values,
these two
processes now become truly complementary, in the sense that
everywhere in the
$(m_A,\tan\beta)$ plane there is a substantial cross-section for at
least one
of them ($\sigma > 0.01$ pb). This should be an excellent starting
point for
experimental searches. Similar considerations hold for $h A$, $H A$
production,
whose cross-sections are shown in fig.~6. As long as one of the two
channels is
kinematically accessible, the inclusive cross-section is large enough
to
provide a  substantial event rate. Even in this case the two
processes are
complementary, and together should be able to probe the region of
parameter
space corresponding to $m_A \simlt 200$ GeV. At a high-energy linear
$\epem$
collider one can also consider single Higgs production via
vector-boson fusion.
The cross-sections for $h$, $H$ production via $WW$ fusion can exceed
0.01~pb
in large, complementary regions of the $\matb$ plane. The $ZZ$ fusion
processes
are suppressed by an order of magnitude with respect to the $WW$
fusion ones,
but could still be useful for experimental searches. Obviously, since
the $AWW$
and $AZZ$ vertices are absent at tree level, one cannot have a
substantial $A$
production with this mechanism for sensible values of the parameters.

The global picture that emerges from these results is the following.
If
no neutral Higgs boson is discovered until then, one must find, at a
linear
$\epem$ collider with $\sqrt{s} = 500 \gev$ (EE500) at least one
neutral SUSY
Higgs, otherwise the MSSM is ruled out (the same applies to its most
plausible
non-minimal extensions). If $m_A$ is not too large, at EE500 there is
the
possibility of discovering all of the Higgs states of the MSSM via a
variety of
processes, including charged-Higgs-boson production, which has not
been
discussed here. In the event that a neutral Higgs boson is already
discovered
at LEP or the LHC/SSC, with properties compatible with one of the
MSSM Higgs states, EE500 would still be a very useful instrument to
investigate
in detail the spectroscopy of the Higgs sector, for example to
distinguish
between the SM, the MSSM and possibly other non-minimal
supersymmetric
extensions.

\section{Conclusions}

In conclusion, the search for Higgs bosons in the low and
intermediate mass
range is a crucial test of the MSSM, and more generally of the whole
idea of
low-energy supersymmetry. Compared with direct searches for $R$-odd
supersymmetric particles, this test has a smaller dependence on
subjective
naturalness bounds on the amount of SUSY breaking. Under the generic
assumption
of $R$-odd particles not much above the TeV scale, the MSSM would be
ruled out
by the experimental exclusion of a Higgs boson below 130~GeV or so.
Non-minimal
models with extra gauge singlets could only survive up to Higgs
masses of 140
GeV or so, under the only extra assumption that dimensionless
couplings do not
blow up below the grand-unification scale. More complicated
non-minimal
extensions could in principle escape the latter bound, but in that
case one
would need very artificial constructions to avoid the constraints
coming from
precision electroweak data and to preserve the successful predictions
of grand
unification. One could then say that ruling out the existence of a
Higgs boson
in the low or intermediate mass range would effectively rule out the
idea of
low-energy SUSY. A more optimistic scenario is the one in which a
SM-like Higgs
boson is indeed found in the low or intermediate mass range. This
could not be
taken as evidence for supersymmetry, but it would certainly give
additional
motivations to expect the existence of other Higgs states and of the
$R$-odd
SUSY
particles at accessible mass scales. Finally, the gold-plated
scenario is the
one in which one finds from the beginning either a Higgs boson with
non-standard properties or some $R$-odd supersymmetric particle: it
is easy to
imagine the theoretical and experimental excitement that such an
event would
generate.

\newpage
\section*{References}
\begin{enumerate}
\item
\label{fayet}
P.~Fayet, Phys. Lett. B69 (1977) 489.
\item
\label{sohnius}
M.F.~Sohnius, Phys. Rep. 128 (1985) 39.
\item
\label{hunter}
J.F.~Gunion, H.E.~Haber, G.L.~Kane and S.~Dawson, {\em The Higgs
Hunter's
Guide} (Addison-Wesley, New York, 1990).
\item
\label {pioneer}
Y.~Okada, M.~Yamaguchi and T.~Yanagida, Prog. Theor. Phys. Lett. 85
(1991) 1;
J.~Ellis, G.~Ridolfi and F.~Zwirner, Phys. Lett. B257 (1991) 83;
H.E.~Haber
and R.~Hempfling, Phys. Rev. Lett. 66 (1991) 1815. Large corrections
to the
neutral Higgs mass sum rule were found by M.~Berger, Phys. Rev. D41
(1990) 225.
\item
\label{lep1}
J.F.~Grivaz, contribution to these proceedings.
\item
\label{lep2}
D.~Treille, contribution to these proceedings.
\item
\label{mnmssm}
H.P.~Nilles, M.~Srednicki and D.~Wyler, Phys. Lett. B120 (1983) 346;
J.M.~Fr\`ere, D.R.T.~Jones and S.~Raby, Nucl. Phys. B222 (1983) 11;
J.-P.~Derendinger and C.A.~Savoy, Nucl. Phys. B237 (1984) 307.
\item
\label{gang}
J.~Ellis, J.F.~Gunion, H.E.~Haber, L.~Roszkowski and F.~Zwirner,
Phys. Rev. D39
(1989) 844.
\item
\label{muproblem}
J.E.~Kim and H.-P.~Nilles, Phys. Lett. B138 (1984) 150.
\item
\label{lepsusy}
R.~Van~Kooten, contribution to these proceedings.
\item
\label{missing}
B.~Grinstein, Nucl. Phys. B206 (1982) 387; A.~Masiero,
D.V.~Nanopoulos,
K.~Tamvakis and T.~Yanagida, Phys. Lett. B115 (1982) 380;
I.~Antoniadis,
J.~Ellis, J.S.~Hagelin and D.V.~Nanopoulos, Phys. Lett. B194 (1987)
231.
\item
\label{pseudo}
K.~Inoue, A.~Kakuto and T.~Takano, Progr. Theor. Phys. 75 (1986) 664;
A.A.~Anselm and A.A.~Johansen, Phys. Lett. B200 (1988) 331;
R.~Barbieri,
G.~Dvali and A.~Strumia, Nucl. Phys. B391 (1993) 487.
\item
\label{giudice}
G.F.~Giudice and A.~Masiero, Phys. Lett. B206 (1988) 480; K.~Inoue,
M.~Kawasaki, M.~Yamaguchi and T.~Yanagida,  Phys. Rev. D45 (1992)
329;
I.~Antoniadis, C.~Mu\~noz and M.~Quir\'os, Ecole Polytechnique
preprint
CPTH-A206.1192, FTUAM~92/35, IEM-FT-63/92; V.S.~Kaplunovsky and
J.~Louis,
preprint CERN-TH.6809/93, UTTG-05-93; J.A.~Casas and C.~Mu\~noz,
preprint
CERN-TH.6764/92, IEM-FT-66/92, FTUAM 92/45.
\item
\label{flat}
R.~Barbieri, S.~Ferrara and C.A.~Savoy, Phys. Lett. B119 (1982) 343;
A.H.~Chamseddine, R.~Arnowitt and P.~Nath, Phys. Rev. Lett. 49 (1982)
970.
\item
\label{sliding}
H.-P.~Nilles, M.~Srednicki and D.~Wyler, Phys. Lett. B124 (1982) 337;
A.B.~Lahanas, Phys. Lett. B124 (1982) 341; A.~Sen, Phys. Rev. D30
(1984) 2608
and D31 (1985) 411.
\item
\label{rga}
R.~Barbieri, M.~Frigeni and M.~Caravaglios, Phys. Lett. B258 (1991)
167;
Y.~Okada, M.~Yamaguchi and T.~Yanagida, Phys. Lett. B262 (1991) 54;
J.R.~Espinosa and M. Quir\'os, Phys. Lett. B266 (1991) 389; M.A. Diaz
and
H.E.~Haber, Phys. Rev. D45 (1992) 4246; K.~Sasaki, M.~Carena and
C.E.M.~Wagner,
Nucl. Phys. B381 (1992) 66; P.H.~Chankowski, S.~Pokorski and
J.~Rosiek, Phys.
Lett. B281 (1992) 100; L.A.C.P.~da Mota, Oxford preprint OUTP-92-31P.
\item
\label{epa}
R.~Barbieri and M.~Frigeni, Phys. Lett. B258 (1991) 395; J.~Ellis,
G.~Ridolfi
and F.~Zwirner, Phys. Lett. B262 (1991) 477; A.~Brignole, J.~Ellis,
G.~Ridolfi
and F.~Zwirner, Phys. Lett. B271 (1991) 123 and (E) B273 (1991) 550;
M.~Drees
and N.M.~Nojiri, Phys. Rev. D45 (1992) 2482; D.M.~Pierce,
A.~Papadopoulos and
S.B.~Johnson, Phys. Rev. Lett. 68 (1992) 3678; S.~Kelley, J.L.~Lopez,
D.V.~Nanopoulos, H.~Pois and K.~Yuan, Phys. Lett. B285 (1992) 61;
H.E.~Haber,
R.~Hempfling and Y.~Nir, Phys. Rev. D46 (1992) 3015; H.E.~Haber and
R.~Hempfling, Santa Cruz preprint SCIPP-91/33; V.~Barger,
M.S.~Berger,
A.L.~Stange and R.J.N.~Phillips, Phys. Rev. D45 (1992) 4128.
\item
\label{da}
A.~Yamada, Phys. Lett. B263 (1991) 233; P.H.~Chankowski, S.~Pokorski
and
J.~Rosiek, Phys. Lett. B274 (1992) 191 and B286 (1992) 307;
A.~Brignole, Phys.
Lett. B277 (1992) 313 and B281 (1992) 284; D.~Pierce and A.
Papadopoulos, Phys.
Rev. D47 (1992) 222.
\item
\label{bz}
A.~Brignole and F.~Zwirner, Phys. Lett. B299 (1993) 72.
\item
\label{kz}
Z.~Kunszt and F.~Zwirner, Nucl. Phys. B285 (1992) 3.
\item
\label{landau}
L.~Durand and J.~Lopez, Phys. Lett. B217 (1989) 463 and Phys. Rev.
D40 (1989)
207; M.~Drees, Int. J. Mod. Phys. A4 (1989) 3635; P.~Bin\'etruy and
C.A.~Savoy,
Phys. Lett. B277 (1992) 453; J.R.~Espinosa and M.~Quir\'os, Phys.
Lett. B279
(1992) 92 and B302 (1993) 51; U.~Ellwanger and M.~Rausch de
Traubenberg,
Z.~Phys. C53 (1992) 521; T.~Moroi and Y.~Okada, Phys. Lett. B295
(1992) 73;
G.~Kane, C.~Kolda and J.~Wells, Michigan preprint UM-TH-92-24; W.~ter
Veldhuis,
Purdue preprint PURD-TH-92-11; T.~Elliott, S.F.~King and P.L.~White,
Southampton preprint SHEP-92/93-11; D.~Comelli, Trieste preprint
UTS-DFT-92-30;
U.~Ellwanger and M.~Lindner, Phys. Lett. B301 (1993) 365;
U.~Ellwanger,
Heidelberg preprint HD-THEP-93-4.
\item
\label{cmpp}
N.~Cabibbo, L.~Maiani, G.~Parisi and R.~Petronzio, Nucl. Phys. B158
(1979) 295.
\item
\label{eq}
J.R.~Espinosa and M.~Quir\'os, Madrid preprint IEM-FT-60-92.
\item
\label{janot}
P.~Janot, in  `Proceedings of the Workshop on $e^+ e^-$ Collisions at
500 GeV:
the Physics Potential' (P.M.~Zerwas ed.), DESY 92-123, p.~107; Orsay
preprint
LAL-92-27.
\item
\label{others}
Z.~Kunszt and F.~Zwirner, in Proceedings of the Large Hadron Collider
Workshop,
Aachen, 1990 (G. Jarlskog and D.~Rein, eds.), Vol.~II, p.~578;
V.~Barger,
M.S.~Berger, A.L.~Stange and R.J.N.~Phillips, Phys. Rev. D45 (1992)
4128;
H.~Baer, M.~Bisset, C.~Kao and X.~Tata, Phys. Rev. D46 (1992) 1067;
J.F.~Gunion
and L.~Orr, Phys. Rev. D46 (1992) 2052; J.F.~Gunion, R.~Bork,
H.E.~Haber and
A.~Seiden, Phys. Rev. D46 (1992) 2040; J.F.~Gunion, H.E.~Haber and
C.~Kao,
Phys. Rev. D46 (1992) 2907; H.~Baer, C.~Kao and X.~Tata, Florida
preprint
FSU-HEP-920717; H.~Baer, M.~Bisset, D.~Dicus, C.~Kao and X.~Tata,
Phys. Rev.
D47 (1993) 1062; V. Barger, K. Cheung, R.J.N. Phillips and
A.L.~Stange, Phys.
Rev. D46 (1992) 4914.
\\
See also the Letters of Intent of the SDC, GEM, ATLAS,
CMS and L3P Collaborations, and the contributions to these
proceedings by
K.~McFarlane (GEM), M.~Mangano (SDC), S.~Hellman (ATLAS), C.~Seez
(CMS) and
F.~Nessi-Tebaldi (L3P).
\item
\label{virtual}
R.~Barbieri and E.~Lisi, contributions to these proceedings.
\item
\label{bsga}
J.L.~Hewett, Phys. Rev. Lett. 70 (1993) 1045; V.~Barger, M.S.~Berger
and
R.J.N.~Phillips, Phys. Rev. Lett. 70 (1993) 1368; T.~Hayashi,
M.~Matsuda and
M.~Tanimoto, Kogakkan Univ. preprint KU-01-93; M.A.~Diaz, Vanderbilt
Univ.
preprint VAND-TH-93-2.
\item
\label{bsgb}
S.~Bertolini, F.~Borzumati and A.~Masiero, Nucl. Phys. B294 (1987)
321;
S.~Bertolini, F.~Borzumati, A.~Masiero and G.~Ridolfi, Nucl. Phys.
B353 (1991)
591; N.~Oshimo, Lisbon preprint CFMC-IFM/92-12; R.~Barbieri and
G.F.~Giudice,
preprint CERN-TH.6830/93.
\item
\label{ee500}
A.~Brignole, J.~Ellis, J.F.~Gunion, M.~Guzzo, F.~Olness, G.~Ridolfi,
L.~Roszkowski and F.~Zwirner, in  `Proceedings of the Workshop on
$e^+ e^-$
Collisions at 500 GeV: the Physics Potential' (P.M.~Zerwas ed.), DESY
92-123,
p.~613.
\item
\label{othersee}
J.F.~Gunion, L.~Roszkowski, A.~Turski, H.E.~Haber, G.~Gamberini,
B.~Kayser,
S.F.~Novaes, F.~Olness and J.~Wudka, Phys. Rev. D38 (1988) 3444;
A.~Yamada,
Mod. Phys. Lett. A7 (1992) 2877; A.~Djouadi, J.~Kalinowski and
P.M.~Zerwas, in
`Proceedings of the Workshop on $e^+ e^-$ Collisions at 500 GeV: the
Physics
Potential' (P.M.~Zerwas ed.), DESY 92-123, p.~83.
\end{enumerate}
\newpage
\begin{figure}
\vspace{10.0cm}
\caption{Contour curves of $m_h^{max}$ in the $(m_t,\msq)$ plane
(from ref.~[20]).}
\end{figure}
\newpage
\begin{figure}
\vspace{10.0cm}
\caption{$\Gamma(H \to h h)$ as a function of $m_H$, for the
indicated
parameter choice. The solid line corresponds to the full diagrammatic
calculation, the dashed line to the effective potential approach, the
dash-dotted line to the `improved tree-level' result (from
ref.~[19]).}
\end{figure}
\newpage
\begin{figure}
\vspace{15.0cm}
\caption{Upper bound on the mass of lightest neutral Higgs boson, as
a function
of $m_t$, in the non-minimal supersymmetric extension of the SM with
an extra
singlet. The different solid lines correspond to the indicated values
of $\tb$,
and all soft SUSY-breaking masses have been chosen equal to the
common value
$M_{\rm SUSY}=1 \tev$ (from ref.~[23]).}
\end{figure}
\newpage
\begin{figure}
\vspace{15.0cm}
\caption{Pictorial summary of the discovery potential of the LHC in
the $(m_A,
\tan \beta)$ plane (from ref.~[20]).}
\end{figure}
\newpage
\begin{figure}
\vspace{21.0cm}
\caption{Contours of a) $\sigma ( e^+ e^- \rightarrow h Z)$ and b)
$\sigma (
e^+ e^- \rightarrow H Z)$, in the $(m_A, \tan \beta)$ plane (from
ref.~[29]).}
\end{figure}
\newpage
\begin{figure}
\vspace{21.0cm}
\caption{Contours of a) $\sigma ( e^+ e^- \rightarrow h A)$ and b)
$\sigma (
e^+ e^- \rightarrow H A)$, in the $(m_A, \tan \beta)$ plane (from
ref.~[29]).}
\end{figure}
\end{document}